\documentstyle[aps,pra,epsf,twocolumn]{revtex}

\begin{document}
\draft

\title{Characterization of a high-power tapered semiconductor amplifier system}

\author{D. Voigt, E.C. Schilder, R.J.C. Spreeuw, and H.B. van Linden van den Heuvell}

\address{Van der Waals-Zeeman Instituut, Universiteit van Amsterdam,\\
         Valckenierstraat 65, 1018 XE Amsterdam, the Netherlands\\
         e-mail: voigt@wins.uva.nl}

\date{\today}

\maketitle

\begin{abstract}

We have characterized a semiconductor amplifier laser system which provides up
to 200~mW output after a single-mode optical fiber at 780~nm wavelength. The
system is based on a tapered semiconductor gain element, which amplifies the
output of a narrow-linewidth diode laser. Gain and saturation are discussed as
a function of operating temperature and injection current. The spectral
properties of the amplifier are investigated with a grating spectrometer.
Amplified spontaneous emission (ASE) causes a spectral background with a width
of 4~nm FWHM. The ASE background was suppressed to below our detection limit by
a proper choice of operating current and temperature, and by sending the light
through a single-mode optical fiber. The final ASE spectral density was less
than 0.1~nW/MHz, i.e. less than 0.2\,\% of the optical power. Related to an
optical transition linewidth of $\Gamma/2\pi=6$~MHz for rubidium, this gives a
background suppression of better than $-82$~dB. An indication of the beam
quality is provided by the fiber coupling efficiency up to 59\,\%. The
application of the amplifier system as a laser source for atom optical
experiments is discussed.

\end{abstract}


\section{Introduction}

The techniques of laser cooling and trapping of neutral atoms
\cite{ChuCohPhi98} require stable, narrow linewidth and frequency tunable laser
sources. Commonly used systems for the near infrared wavelengths are based on
external grating diode lasers (EGDL) \cite{WieHol91}. Optical feedback from a
grating narrows the linewidth to less than 1~MHz and provides tunability.
High-power single transverse mode diode lasers can provide up to 80~mW optical
output at wavelengths below 800~nm. In this power range diode lasers thus
provide a less costly alternative to {Ti:Sapphire} lasers.

If more power is required, the output of an EGDL can be amplified. Presently,
there are three common techniques based on semiconductor gain elements:
(i) Injection-locking of a single-mode laser diode \cite{Had86} by seeding
light from an EGDL results typically in 60-80~mW optical power at 780~nm
wavelength.
(ii) Amplification in a double-pass through a broad-area emitting diode laser
(BAL) \cite{AbbYanCha88,GolMehSur93,FeyBeiLee97,BALIgor}. This yields an
optical output of typically 150~mW after spatial filtering. A disadvantage is
the relatively low gain of 10-15, requiring high seed input power. The BAL gain
can be improved using phase conjugating mirrors in the seed incoupling setup
\cite{IidHorMat98}.
(iii) Travelling-wave amplification in a semiconductor gain element with a
tapered waveguide (TA) \cite{GolMehSur93,Wal96,WalKinChi92}. Compared to a BAL
this yields higher gain and higher power after spatial filtering. This approach
requires much lower input and a less complex optical setup than a BAL. However,
a TA gain element is considerably more expensive.

In this work we investigate a TA system that amplifies the narrow-linewidth
seed beam of an EGDL and provides up to 200~mW optical output from a
single-mode optical fiber. We operate the system on the
$D_2$ ($5S_{1/2}\rightarrow 5P_{3/2}$) transition of rubidium, at a wavelength
of 780~nm. The input facet of the tapered gain element element has the typical
width of a low power single-transverse mode diode laser ($\approx 5~\mu$m). A
seed beam is amplified in a single pass and expanded laterally by the taper to
a width of typically $100-200~\mu$m \cite{Wal96} such that the light intensity
at the output facet is kept below the damage threshold and the beam remains
diffraction limited. The output power can be much larger than from a
single-mode waveguide.

In previous work, TA's have been used as sources for frequency-doubling and
pumping solid state lasers \cite{ZimWalHan96}. Apart from the achievable output
power, frequency tunability of the narrow-linewidth output \cite{WanLasAlv98},
simultaneous multifrequency generation \cite{FerMewSch99}, and spatial mode
properties, including coupling to optical fibers \cite{LivChiKin94} have been
addressed.

In this paper we investigated the broadband spectral properties of the TA. The
background due to amplified spontaneous emission (ASE) \cite{ChoCrai90} in the
gain element was minimized by adjusting the amplifier's operating conditions,
{i.e.}~temperature, injection current and seed input power. We also
investigated the coupling efficiency of the TA output to a single-mode optical
fiber, and find that the latter acts both as {\em spatial} and {\em spectral}
filter. The properties of three gain elements of same type are compared.

Atom optical applications usually require good suppression of spectral
background. {E.g.}~in far off-resonance optical dipole traps
\cite{GriWeiOvc00}, scattering of resonant light from the background causes
heating and atom loss. We discuss the consequences of ASE background in such
schemes.

\section{Amplifier setup}

\begin{figure}
 \centerline{\epsfxsize=8.5cm\epsffile{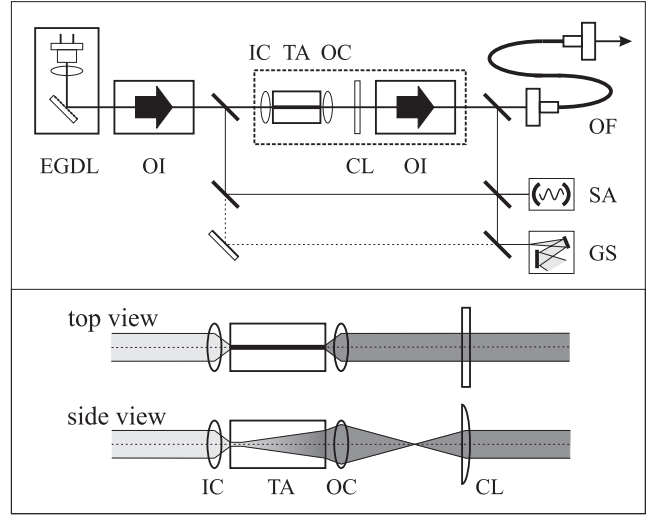}}
 \vspace*{0.5cm}
 \caption{Amplifier setup with seed laser and gain element:
	  external grating diode laser (EGDL),
	  60~dB optical isolators (OI),
	  tapered amplifier (TA),
	  single-mode optical fiber (OF),
	  optical spectrum analyzer (SA), and
	  grating spectrometer (GS).
	  A top and side view of the gain element is shown with
	  input and output collimators (IC,OC),
          and a cylindrical lens (CL) compensating for astigmatism
	  (not to scale).}
 \label{fig:TAsetup}
\end{figure}

Our amplifier system consists of a seed laser, the output beam of which is
amplified in a single pass by the tapered gain element, as shown in
Fig.\,\ref{fig:TAsetup}. The TA output is coupled into a single-mode optical
fiber \cite{OFR}. The seed laser is an EGDL with a linewidth of less than
1~MHz. It contains a 60~mW single-mode laser diode ({\em Hitachi},
{HL\,7851\,G98}). From the EGDL we have 28~mW of power available to seed the
amplifier at 780~nm wavelength. Coupling of the seeding beam to the amplifier
is done by mode-matching the seed laser with the backwards travelling beam
emitted by the TA. The divergence angles from seed laser emission and backward
directed TA emission are similar. Hence, sufficient mode-matching is obtained
using identical collimation lenses for both ({$f=4.5$~mm}, {N.A.=0.55}).
Additional mode shaping, e.g.~with anamorphic prism pairs is not necessary. An
optical isolator with 60~dB isolation protects the stabilized seed laser from
feedback by the mode-matched beam of the amplifier. The 5~mm aperture of the
isolator is sufficiently large that it does not clip the elliptical seed beam.

The TA was a {SDL\,8630\,E} ({ser.no.}~{TD\,310} \cite{SDL}). According to the
manufacturer's data sheet the output power ranges from $0.5-0.55$~W within a
wavelength tuning range from $787-797$~nm, at an operating temperature of
{$21^{\circ}$\,C}. The beam quality parameter is specified as typically
$M^2<1.4$ \cite{SDL}. The TA should be protected from any reflected light,
because it will be amplified in the backward direction and may destroy the
entrance facet. We used an output collimator of large numerical aperture
({$f=3.1~$mm}, {N.A.=0.68}) and sent the beam through a second 60~dB optical
isolator. The plane of the tapered gain element is vertically oriented, so that
diffraction yields a large horizontal divergence. This is collimated similar to
the seed input, but yields a focus in the vertical plane. With a cylindrical
lens ($f=100$~mm), we compensated for astigmatism of the beam, in order to
couple into a single-mode optical fiber. The astigmatism correction is also
shown in Fig.\,\ref{fig:TAsetup}.

There is a considerable loss in optical power due to the isolator transmission.
Taking also into account small reflection losses of the lens surfaces, we
estimate the useful output power to be 78\,\% of the optical power emitted by
the TA facet. In the remainder of this paper, all quoted powers are as measured
with a power meter behind the optical isolator.

The amplifier was provided as an open heatsink device. We mounted it on a water
cooled base and stabilized it to the desired operating temperature within a few
mK by a 40~W thermo-electric cooler. Thermal isolation from the ambient air and
electromagnetic shielding were provided by a metal housing. When operating the
amplifier at temperatures below the dew point, we flushed the containment with
dry nitrogen.

It is necessary to have a compact, stable mounting of the gain element and
collimators. We mounted the collimators in a commercial $xy$ flexure mount to
allow for lateral lens adjustment. The axial $z$ adjustment is done by two
small translation stages. All adjustments except the $z$ direction of the
output collimator are accessible from outside. This proved to be very
convenient for mode-matching the seed beam and also for compensating beam
displacement of the TA output when changing temperature or current.  

The narrow spectral line of seed laser and amplifier output was monitored by an
optical spectrum analyzer of 1~GHz free spectral range with 50~MHz resolution.
The amplifier's broad spectral background was analyzed using a grating
spectrometer with a resolution of 0.27~nm. Also the output of the single-mode
fiber was recorded with the spectrometer.

\section{Unseeded operation of the amplifier}

\begin{figure}
 \centerline{\epsfxsize=9.5cm\epsffile{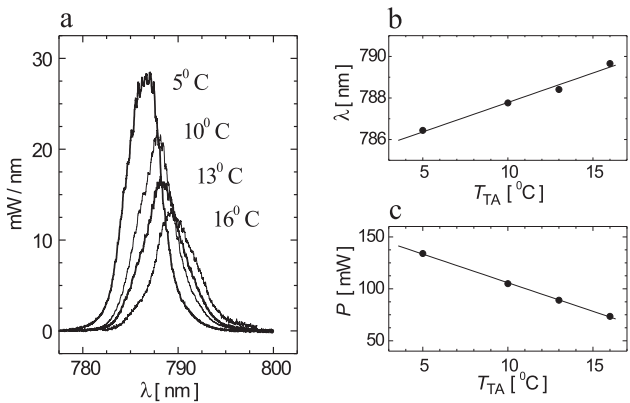}}
%\vspace*{0.5cm}
 \caption{Temperature dependence of the unseeded amplifier at 1.2~A injection
          current:
		(a) spectra,
		(b) center wavelengths,
		(c) output power after optical isolator.
		Solid lines indicate linear fits.}
 \label{fig:FreeRun}
\end{figure}

When the TA receives no seed input, it operates as a laser diode. Thus, when
the injection current, $I_{TA}$, is increased from zero, the optical output
power shows the lasing threshold (see Fig.\,\ref{fig:Amplification}ab).
Generally, both the operating wavelength and the optical power of a laser diode
depend on the temperature. This property is shown in Fig.\,\ref{fig:FreeRun}.
The emission spectrum of the lasing tapered gain element is almost Gaussian
shaped, with an $1/e^2$ width of 4~nm. It appears as a background of ASE also
in the spectra when operating the gain element as an amplifier (see below). The
oscillatory structures on the spectra are artifacts of the spectrometer. In the
fitted spectra, we evaluated the center wavelength at each temperature setting.
It increases with temperature with a slope of {0.28~nm/K}, typical for
semiconductor lasers (Fig.\,\ref{fig:FreeRun}b).

The temperature dependence of the output power is shown in
Fig.\,\ref{fig:FreeRun}c. We operated the TA within the specifications of the
manufacturer's data sheet, that recommends to keep the optical power at the
output facet below 550~mW. As the temperature increases, the conversion
efficiency (mW/A) decreases and the threshold current increases. This can be
seen in Fig.\,\ref{fig:Amplification}ab (open symbols) where we plot the
optical output power $P$ vs. current $I_{TA}$ for two temperature settings. The
threshold current of the unseeded TA increases from 0.78~A ($5^\circ$\,C) to
0.86~A ($14^\circ$\,C). From the slopes above threshold, we find that the
conversion efficiency decreases from 0.7~W/A ($5^\circ$\,C) to 0.5~W/A
($14^\circ$\,C). In order to measure the unperturbed output of the unseeded TA,
one has to prevent light emitted from the entrance facet of being reflected.
Even a very weak reflection would be amplified in the forward direction.

For the unseeded TA, we also measured the light propagating backwards from the
amplifier's entrance facet. It reaches typically a power of $10-25$~mW for
injection currents from $1-1.4$~A. Hence the necessity of a good isolation of
the seeding laser.

\section{Amplification of a seed beam}

\begin{figure}
\centerline{\epsfxsize=9.5cm\epsffile{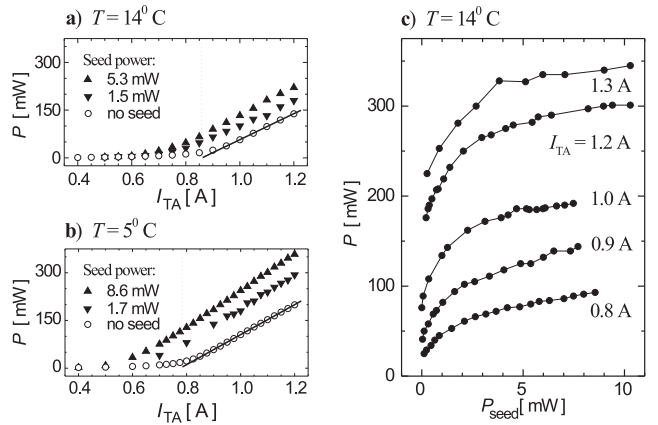}}
%\vspace*{0.5cm}
 \caption{Amplifier output vs.~injection current $I_{TA}$
	  and seed power $P_{seed}$.
	  The lasing thresholds for the unseeded amplifier are
	  0.86~A ($14^\circ$~C) and 0.78~A ($5^\circ$~C),
	  indicated by vertical dotted lines.}
 \label{fig:Amplification}
\end{figure}

Amplification of a seed beam is evident in the output power of the TA. In
Fig.\,\ref{fig:Amplification}ab, we have plotted the output power for distinct
seed powers, $P_{seed}$, at two temperature settings. For the larger seed
inputs of 8.6~mW and 5.3~mW, respectively, the amplifier was well saturated.
The saturation is evident from Fig.\,\ref{fig:Amplification}c where $P_{seed}$
was varied for injection currents from $0.8-1.3$~A.
With $P_{seed}\approx 4$~mW
the device appears to be saturated for all current settings. For seed powers
between $2-4$~mW, the amplification ranges from $70-140$, e.g.~320~mW output
with 4~mW seed.

\begin{figure}
\centerline{\epsfxsize=9.5cm\epsffile{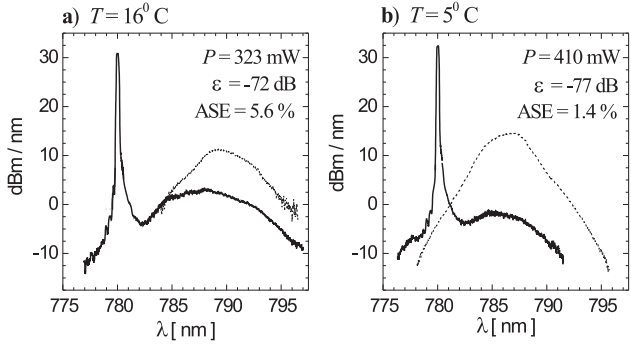}}
%\vspace*{0.5cm}
 \caption{Spectrum of the amplifier output (before the fiber).
	  The seed power is 28~mW at 1.2~A injection current.
	  The dashed curves are for unseeded operation.
	  $P$ is the total optical power,
	  ASE is the fraction of background power,
          and $\epsilon$ is the ASE suppression for the power
	  spectral density in units of mW/$\Gamma$ ($\Gamma/{2\pi}=6$~MHz).}
 \label{fig:SeededSpectrum}
\end{figure}

We now discuss the spectral properties of the TA and in particular the
suppression of ASE background. Fig.\,\ref{fig:SeededSpectrum} shows the power
spectral density of the TA output {\it before} an optical fiber for
$16^\circ$\,C and $5^\circ$\,C operating temperature. In both cases the
amplifier was saturated with 28~mW seed input. For comparison also the
corresponding spectra of the unseeded amplifier are shown. In saturation, the
broad ASE background is distinguished from a narrow peak of the amplified seed
signal. The width of the peak is given by the bandwidth of the spectrometer,
0.27~nm FWHM. (By means of an optical spectrum analyzer and Doppler-free
spectroscopy signals of rubidium, we could verify that the amplified signal has
a narrow width comparable to that of the EGDL.) 

The influence of the operating temperature is obvious first by the increased
output power at lower temperature: 323~mW ($16^\circ$\,C) and 410~mW
($5^\circ$\,C), respectively. Second, both the peak level and total amount of
ASE background are better suppressed at lower temperature. We attribute this to
the shift of the gain profile of the TA towards the seed wavelength of 780~nm
at lower temperature \cite{ChoCrai90}. The fraction of ASE background in the TA
output is obtained by integrating the power spectral densities in
Fig.\,\ref{fig:SeededSpectrum}, yielding 5.6\,\% ($16^\circ$\,C) and 1.4\,\%
($5^\circ$\,C), respectively.

More than the total ASE fraction, for atom-optical applications the important
figure is the fraction of ASE within the natural linewidth of the atomic
transition used. We define this ratio $\epsilon$ by comparing the power in the
peak with the ASE power in a bandwidth given by a typical atomic natural
linewidth, {e.g.}~$\Gamma/{2\pi}=6$~MHz for rubidium. For $16^\circ$\,C
($5^\circ$\,C), the peak value of +2.5~dBm/nm ($-2.0$~dBm/nm) of ASE is then
reexpressed as 22~nW/$\Gamma$ (7.9~nW/$\Gamma$). With 323~mW (410~mW) in the
narrow line, the suppression ratio $\epsilon$ is $-72$~dB ($-77$~dB). We can
thus optimize the spectral properties of the TA output by an appropriate choice
of operating temperature. Even better suppression can be achieved by use of an
optical fiber as {\em spectral} filter, as discussed in the following section.

\begin{figure}
 \centerline{\epsfxsize=9.5cm\epsffile{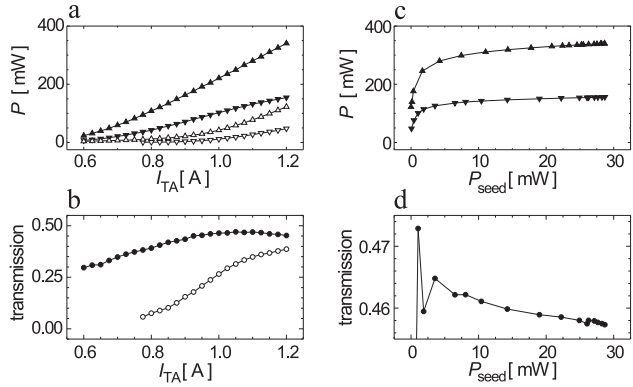}}
%\vspace*{0.5cm}
 \caption{Transmission through a {single-mode} optical fiber.
    (a) Fiber input and output power with and without seed,
    (b) fiber transmission with and without seed,
    (c) Fiber input and output in dependence on seed power,
    (d) corresponding fiber transmission.
	The symbols: seeded with 28~mW (solid), unseeded (open),
                     fiber input (up triangles), fiber output (down triangles),
		     fiber transmission (circles).}
  \label{fig:FiberTransmission}
\end{figure}

\section{Spatial and spectral filtering using an optical fiber}

For many applications, laser beam quality is an important property, e.g.~for
optical dipole traps. A convenient method to obtain spatial filtering is to
send the light through a single-mode optical fiber. An additional advantage of
the fiber is a decoupling of optical alignment between different parts of the
experimental setup. Here the coupling efficiency is discussed and the spectrum
of the transmitted light is compared with the spectrum before the fiber. We
observe that {\em spatial} filtering by the fiber is accompanied by
{\em spectral} filtering. Evidently, the contribution of ASE in the TA beam is
spatially distinguishable from the amplified seed signal.

We find that the spatial mode properties of the saturated TA output are
slightly different for different injection currents.
Fig.\,\ref{fig:FiberTransmission}ab represents the fiber transmission
vs.~current, $I_{TA}$. The fiber coupling was optimized for a current of 1~A
and the TA was saturated. A maximum transmission of 46\,\% is achieved. For
comparison, with an EGDL, after circularizing the beam using an anamorphic
prism pair, we typically obtain a fiber transmission of 75\,\%. The slope in
the transmission curve is probably due to a beam displacement caused by the
current-dependent thermal load of the gain element. Such a displacement was
also observed when the operating temperature was changed. With the fiber
coupling thus optimized, light from the unseeded TA has less transmission than
the amplified seed signal. Fig.\,\ref{fig:FiberTransmission}cd shows for a
fixed current of 1~A, that the fiber transmission is almost independent of the
seed input power, i.e.~the beam shape does not change.

\begin{figure}
 \centerline{\epsfxsize=9.5cm\epsffile{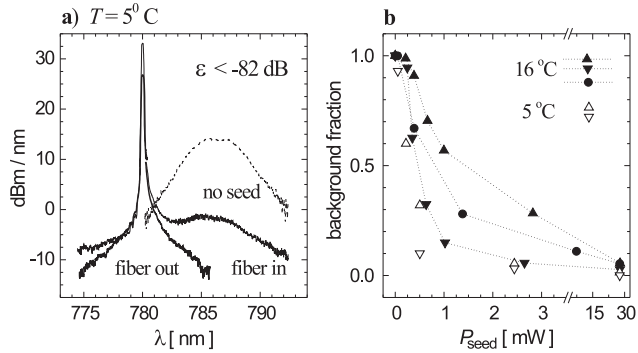}}
%\vspace*{0.5cm}
 \caption{Spectral filtering by a single-mode optical fiber.
    (a) Saturated with 28~mW seed power at 1.2~A current, 130~mW power after
	the fiber, 410~mW in front of it.
	ASE background is not distinguishable from the spectrometer
	response function after the fiber. 
    (b) The ASE fraction depends on the saturation:
	fiber input (up triangles) and output (down triangles)
	at 1.2~A current.
	For comparison: fiber input with 1.45~A current (circles).}
  \label{fig:FiberCoupling}
\end{figure}

Also the light after the fiber has been analyzed using the grating
spectrometer, see Fig.\,\ref{fig:FiberCoupling}a for an operating temperature
of $5^\circ$\,C. For the saturated amplifier a spectral ASE background cannot
be distinguished after the fiber, since the peak is identical with the
spectrometer response function. (This response function was obtained by
recording the spectrum of the narrow-linewidth EGDL laser. A similar response
was also obtained using a {HeNe} laser.) Thus we can only give an upper limit
for the ASE contribution of 0.2\,\%. The suppression ratio is
$\epsilon<-82$~dB, with an ASE level of less than $-12.5$~dBm/nm or
0.7~nW/$\Gamma$, respectively. This should be compared to the value of
$\epsilon=-77$~dB before the fiber, as seen in Fig.\,\ref{fig:SeededSpectrum}b
for $5^\circ$\,C. For comparison, at $16^\circ$\,C, we found an ASE suppression
of only $-76$~dBm {\em after} the fiber. 

The ASE background depends also on the degree of amplifier saturation, as shown
in Fig.\,\ref{fig:FiberCoupling}b. The ASE fraction is plotted vs.~seed power
for light before and after the fiber. It decreases quickly as the TA saturates.
From the spectra acquired before the fiber (up triangles), it is evident that
the increase of seed power into the saturated regime suppresses the ASE.
Whereas mode matching of the seed input beam was not difficult for achieving
maximum output power, optimal ASE suppression required a more careful
alignment, thus optimizing the TA saturation.

It is also obvious from the figure, that larger gain of the TA improves the
output spectrum (circles, larger operating current).

We can summarize the results of Sec.~{IV} and {V} as follows: The spectral
properties of the TA can be optimized by choosing an appropriate operating
temperature, spectral filtering with an optical fiber and saturation of the
gain element.

\section{Variation in the properties of individual gain elements}

We compared the gain element with two other gain elements of the same type
(8630\,E). One gain element (ser.~no.~TD\,430, 777~nm) was used in the setup
described above. A second (ser.~no.~TD\,387, 790~nm) was implemented in a
commercial TA system ({\em TUI Optics GmbH}, TA\,100) and operated on both the
$D_2$ and the $D_1$ ($5S_{1/2}\rightarrow 5P_{1/2}$) transition of rubidium at
780~nm and 795~nm, respectively.

For the different gain elements, we found considerable differences in their
beam quality and consequently their fiber coupling efficiency. Whereas TD\,310
and TD\,387 showed a dominant double-lobed mode structure in the far field and
permitted only a fiber transmission of 46\,\%, the TD\,430 beam showed a less
pronounced lobe structure \cite{TAdestruction}. With this gain element, we
could couple 59\,\% to the fiber and obtained 200~mW of optical power after the
fiber and an ASE suppression of better than $-84$~dBm.

Also the amplification properties showed striking differences among the gain
elements. TD\,430 has similar saturation properties as TD\,310. In contrast,
TD\,387 operates at maximum output power already without seed. This may be due
to different antireflection coatings at the TA facets. Hence, the TD\,387
requires (permanent) monitoring by a spectrometer in order to optimize seed
incoupling and ASE suppression. The current of the TD\,387 cannot be tuned
continuously, because it shows discrete ``locking-ranges'', resembling the
injection-locking behavior of single-mode diode lasers.

\section{Far off-resonance dipole potentials with spectral background}

In this section we present an estimate of the consequences of the broad
spectral ASE background for light scattering in optical dipole traps. A
background that covers atomic resonances, leads to extra resonant scattering.
Usually the detuning $\delta$ for a dipole trap is chosen as large as possible,
given the available laser intensity $I$. The reason is that off-resonance
scattering scales as $\Gamma'_{OR}\propto I/\delta^2$ at low saturation and
large detuning, whereas the dipole potential is only inversely proportional to
the detuning, ${\cal U}\propto I/\delta$ \cite{CohDupGry92}.

In the presence of a resonant background the total scattering rate of the atoms
is $\Gamma'=\Gamma'_{OR}+\Gamma'_{R}$, where $\Gamma'_{R}$ represents the
resonant scattering. For a fixed depth of the optical dipole potential this
results in a maximum useful laser detuning, $\delta_{max}$, at which the
scattering rate of the atoms, $\Gamma'$, is minimized.

With low atomic saturation by a weak spectral background, we can write
$\Gamma'_R\approx (\Gamma\pi/4) \epsilon I/I_0$. Here $I_0$ is the saturation
intensity, i.e.~1.6~mW/cm$^2$ for the $D_2$ line of rubidium. Hence, with the
requirement of a fixed potential ${\cal U}$, the two scattering contributions
scale as $\Gamma'_{OR}\propto 1/\delta$ and $\Gamma'_{R}\propto\delta$,
respectively. This results in the optimum detuning and minimum scattering
rate,
\begin{eqnarray}
\delta_{max} & = & \pm \Gamma/\sqrt{2\pi\epsilon},    \nonumber \\
\Gamma' & = & 2\sqrt{2\pi\epsilon}\,{\cal U}/\hbar.   \nonumber
\end{eqnarray}
As an example we consider atoms cooled to a temperature of a few $\mu$K in
optical molasses and require an optical potential depth of
${\cal U}/h\approx 1$~MHz. If the allowable scattering rate is,
e.g.~$\Gamma_{max}<100$~s$^{-1}$, this yields a required background suppression
$\epsilon<-110$~dB and optimum detuning $\delta_{max}\approx 760$~GHz. Such a
small background contribution is of course beyond the resolution of our
spectrometric data, with which we observe at best an upper limit of
$I_R(\delta)<0.7$~nW$/\Gamma$ for a total optical power of 200~mW. This
corresponds to a background suppression of $\epsilon<-84$~dB. With a detuning
of 760~GHz, the extension of the optical potential is restricted to less than
$250~\mu$m.

\section{Conclusions}

We have investigated a tapered semiconductor amplifier system, that provides
150-200~mW narrow linewidth output from a single-mode optical fiber, where the
fiber transmission is up to 59\,\%, depending on the actual gain element in
use. The system requires less than 5~mW seed input to saturate with an 
amplification up to 140 at this seed level. The output of the amplifier
includes a broad spectral background of amplified spontaneous emission. We have
found three means of reducing this background:
(i) Choosing the operating temperature such that the gain profile of the
amplifier is spectrally centered as close as possible to the amplified
wavelength, (ii) spectrally filtering the output beam with a single-mode
optical fiber, and (iii) saturating the amplifier with sufficient seed input
power. With those measures, the ASE background is below the resolution of our
spectrometer. That is, the ASE fraction is less
than 0.2\,\% of the optical power in the beam and the peak level is less than
0.1~nW/MHz. Relating the power spectral density of the background to the
natural transition linewidth of rubidium ($\Gamma/{2\pi}=6$~MHz), the ASE
suppression is better than -82~dB. 

We discussed the atom optical application of such an amplifier system with far
off-resonance dipole potentials. A broad ASE background implies here an optimum
laser detuning with which light scattering by atoms is minimized.

A tapered amplifier system may be a lower-cost option to a {Ti:Sapphire} laser.
The available single-transverse mode optical power and spectral properties are
similar to those of broad-area semiconductor laser amplifiers.

\section{Acknowledgements}

We wish to thank K. Dieckmann, A. G{\"o}rlitz, W. Kaenders, J. Schuster,
I. Shvarchuck, B. Wolfring, A. Zach, and C. Zimmermann for helpful information.
This work is part of the research program of the ``Stichting voor Fundamenteel
Onderzoek van de Materie'' (Foundation for the Fundamental Research on Matter)
and was made possible by financial support from the ``Nederlandse Organisatie
voor Wetenschappelijk Onderzoek'' (Netherlands Organization for the Advancement
of Research). R.S. has been financially supported by the Royal Netherlands
Academy of Arts and Sciences.

\bibliographystyle{prsty}

\end{document}